\documentclass{ws-procs9x6}

\usepackage{psfrag}

\psfrag{Q2}{{\tiny $Q^2$}}

\psfrag{G1pGD}{{\scriptsize $G^{\pi^0 p}_1/G_D$}}

\psfrag{G2pGD}{{\scriptsize $G^{\pi^0 p}_2/G_D$}}

\psfrag{G1nGD}{{\scriptsize $G^{\pi^+ n}_1/G_D$}}

\psfrag{G2nGD}{{\scriptsize $G^{\pi^+ n}_2/G_D$}}

\psfrag{EpGd}{{\scriptsize $E^{\pi^0 p}_{0+}/G_D$}}

\psfrag{LpGd}{{\scriptsize $L^{\pi^0 p}_{0+}/G_D$}}

\psfrag{EnGd}{{\scriptsize $E^{\pi^+ n}_{0+}/G_D$}}

\psfrag{LnGd}{{\scriptsize $L^{\pi^+ n}_{0+}/G_D$}}

\psfrag{pin}{{\scriptsize $\pi^+n$}}

\psfrag{pip}{{\scriptsize $\pi^0p$}}

\psfrag{pin}{{\scriptsize $\pi^+n$}}

\psfrag{pip}{{\scriptsize $\pi^0p$}}

\psfrag{G1GA}{{\scriptsize $\sqrt{2}Q^2G_1^{\pi N}/(G_A m_N^2)$}}

\psfrag{G2GA}{{\scriptsize $\sqrt{2}|G_2^{\pi N}|/G_A$}}

\psfrag{F2}{{\scriptsize $10^3\times F_2^p(W,Q^2)$}}

\psfrag{W2}{{\scriptsize $W^2$,~GeV$^2$}}

\psfrag{pi0piplus}{{\scriptsize $F_2^{\gamma^*p\to \pi^0 p}/F_2^{\gamma^*p\to X}$}}

\psfrag{dsigma}{{\scriptsize $Q^6 \sigma_{\gamma^*p\to \pi^0 p}$}}

\psfrag{ds}{{\footnotesize $d\sigma_{\gamma^*p\to \pi^0 p}/d\Omega_\pi,~\mu\mbox{b/ster}$}}

\psfrag{cost}{{\footnotesize $\cos\theta$}}

\newcommand{\beq}[1]{
\begin{equation}\label{#1}}
\newcommand{\eeq}{\end{equation}}
\newcommand{\bea}[1]{
\begin{eqnarray}\label{#1}}
\newcommand{\eea}{\end{eqnarray}}

\begin{document}

\title{ELECTROPRODUCTION OF SOFT PIONS \\ AT LARGE MOMENTUM TRANSFERS}

\author{V. M. BRAUN$^1$,  D. Yu. IVANOV$^2$ and A. PETERS$^1$}

\address{{}$^1$ Institut f\"ur Theoretische Physik, Universit\"at
          Regensburg, \\ D-93040 Regensburg, Germany\\
  {}$^2$ Sobolev Institute of Mathematics, 630090 Novosibirsk, Russia}

\begin{abstract}
We consider pion electroproduction on a proton target
close to threshold for $Q^2$ in the region $1-10$~GeV$^2$.
The momentum transfer dependence of the 
S-wave multipoles at threshold, $E_{0+}$ and $L_{0+}$, is calculated 
using light-cone sum rules. 
\end{abstract}


\bodymatter

\section{Threshold Pion Production}\label{sec1}
Pion electroproduction at threshold from a proton target
\bea{piprod}
  e(l)+p(P) &\to& e(l') + \pi^+(k) + n(P')\,,
\nonumber\\
  e(l)+p(P) &\to& e(l') + \pi^0(k) + p(P')\,
\eea
can be described in terms of two generalised form factors defined as \cite{Braun:2006td}
\bea{def:G12}
\lefteqn{
 \langle N(P')\pi(k) |j_\mu^{em}(0)| p(P)\rangle=}
\\ &=& 
- \frac{i}{f_\pi} \bar N(P')\gamma_5
  \left\{\left(\gamma_\mu q^2 - q_\mu \!\not\! q\right) \frac{1}{m^2} G_1^{\pi N}(Q^2)
    - \frac{i \sigma_{\mu\nu}q^\nu}{2m} G_2^{\pi N}(Q^2)\right\}N(P)\,,
\nonumber
\eea
which can be related to the S-wave transverse $E_{0+}$ and longitudinal $L_{0+}$ multipoles:
\begin{eqnarray}
E_{0+}^{\pi N}&=&\frac{\sqrt{4\pi \alpha_{\rm em}}}{8\pi f_{\pi}}
\sqrt{\frac{(2m+m_{\pi})^2+Q^2}{m^3(m+m_\pi)^3}}\left(Q^2 G_1^{\pi N}-\frac12 {m m_\pi} G_2^{\pi N}\right),
\nonumber\\
L_{0+}^{\pi N}&=&\frac{\sqrt{4\pi \alpha_{\rm em}}}{8\pi f_{\pi}}
\frac{m|\omega^{\rm th}_\gamma|}{2}
\sqrt{\frac{(2m+m_\pi)^2+Q^2}{m^3(m+m_{\pi})^3}}\left(G_2^{\pi
N}+\frac{2m_\pi}{m}G_1^{\pi N}\right).
\label{waves}
\end{eqnarray}
The differential cross section at threshold is given by
\begin{equation}
 \frac{d\sigma_{\gamma^*}}{d\Omega_\pi}\Big|_{\rm th} = \frac{2|\vec{k}_f| W}{W^2-m^2}
\Big[  (E^{\pi N}_{0+})^2 + \epsilon \frac{Q^2}{(\omega^{\rm th}_\gamma)^2}
(L^{\pi N}_{0+})^2\Big].
\label{sigma_th}
\end{equation}
Here and below $m=939$~MeV is the nucleon mass, $W^2=(k+P')^2$ is the invariant energy, 
$\vec{k}_f$ and $\omega^{\rm th}_\gamma$ are the pion three-momentum and the photon energy in the c.m. frame. 
The generalised form factors in (\ref{def:G12}) are real functions of the momentum transfer $Q^2$ at the threshold
$W_{\rm} = m+m_\pi$. For generic $W$ the definition in (\ref{def:G12}) can be extended to specify two of the 
existing six invariant amplitudes, $G_{1,2}(Q^2)\to G_{1,2}(Q^2,W)$, which become complex functions.  

The celebrated low-energy theorem (LET) \cite{KR,Nambu:1997wa,Nambu:1997wb} relates the S-wave multipoles or, equivalently,
the form factors $G_1,G_2$ at threshold, to the nucleon electromagnetic and axial form factors for vanishing pion 
mass $m_\pi=0$
\begin{eqnarray}
 \frac{Q^2}{m^2} G_1^{\pi^0 p} &=& \frac{g_A}{2}\frac{Q^2}{(Q^2+2m^2)} G_M^p\,,
\label{LET}
\quad  G_2^{\pi^0 p} =  \frac{2 g_A m^2}{(Q^2+2m^2)} G_E^p\,,
\\
  \frac{Q^2}{m^2} G_1^{\pi^+ n} &=& \frac{g_A}{\sqrt{2}} \frac{Q^2}{(Q^2+2m^2)} G_M^n + \frac{1}{\sqrt{2}}G_A\,,
 \quad G_2^{\pi^+ n} =   \frac{2\sqrt{2} g_A m^2}{(Q^2+2m^2)} G_E^n\,.
\nonumber
\end{eqnarray} 
Here the terms in $G_{M,E}$ are due to pion emission off the initial proton state, 
whereas for charged pion in addition there is a contribution corresponding
to the chiral rotation of the electromagnetic current. 

The subsequent discussion concentrated mainly on the corrections to (\ref{LET}) due to finite pion mass 
\cite{Vainshtein:1972ih,Scherer:1991cy}.
More recently, the threshold pion production for small $Q^2$ was reconsidered and the low-energy theorems re-derived 
in the framework of the chiral perturbation theory (CHPT), see \cite{Bernard:1995dp} for a review. The new insight gained from 
CHPT calculations \cite{Bernard:1992ys} is that the expansion at small $Q^2$ has to be done with care as the limits 
$m_\pi\to0$ and $Q^2\to 0$ do not commute, in general. The LET predictions seem to be in good agreement with experimental 
data on pion photoproduction \cite{Drechsel:1992pn}, However, it appears \cite{Bernard:1992rf,Bernard:1995dp} 
that the S-wave electroproduction cross section (\ref{sigma_th}) for already 
$Q^2 \sim 0.1$~GeV$^2$ cannot be explained without taking into account chiral loops.   
    
{}For larger momentum transfers the situation is much less studied as the power counting of CHPT cannot be applied.
The traditional derivation of LET using PCAC and current algebra does not seem to be affected as long as
the emitted pion is  'soft' with respect to the initial and final state nucleons simultaneously.
The corresponding condition is, parametrically, $Q^2 \ll \Lambda^3/m_\pi$ (see, e.g. \cite{Vainshtein:1972ih})
where $\Lambda$ is some hadronic scale, and might be satisfied for $Q^2\sim 1$~GeV$^2$ or even higher.  
We are not aware of any dedicated analysis of the threshold production in the $Q^2\sim 1$~GeV$^2$ region, 
however.

{}It was suggested in Ref.~\cite{PPS01} that in the opposite limit of very large momentum transfers 
the standard pQCD collinear factorisation approach \cite{Efremov:1979qk,Lepage:1980fj}
becomes applicable and the helicity-conserving $G_1^{\pi N}$ form factor can be calculated for $m_\pi=0$ 
in terms of chirally rotated nucleon distribution amplitudes. 
In practice one expects that the onset of the pQCD regime is postponed
to very large momentum transfers because the factorisable contribution involves a small factor $\alpha_s^2(Q)/\pi^2$ 
and has to win over nonperturbative ``soft'' contributions that are suppressed by an extra power of $Q^2$ but do not 
involve small coefficients.    

The purpose of this study is to suggest a realistic QCD-motivated model for the $Q^2$ dependence of the $G_{1,2}$ 
form factors alias S-wave multipoles at threshold in the region $Q^2 \sim 1-10$~GeV$^2$ that can be accessible
in current and future experiments in Jefferson Laboratory and elsewhere (HERMES, MAMI).    

\section{Light-Cone-Sum Rules}

In Ref.~\cite{Braun:2001tj} we have developed a technique  to calculate baryon form factors
for moderately large $Q^2$ using light-cone  sum rules (LCSR) \cite{Balitsky:1989ry,Chernyak:1990ag}.  
This approach is attractive because in LCSRs  ``soft'' contributions to the form factors are calculated in 
terms of the same nucleon distribution amplitudes (DAs) that
enter the pQCD calculation and there is no double counting. Thus, the LCSRs provide one with the most 
direct relation of the hadron form factors and distribution amplitudes that is available at present, 
with no other nonperturbative parameters. 

The same technique can be applied to pion electroproduction. 
In Ref.~\cite{Braun:2006td} the $G_1$ and  $G_2$ form factors were estimated in the LCSR approach for 
the range of momentum transfers $Q^2 \sim 5-10$~GeV$^2$. 
{}For this work,  we have reanalysed the sum rules derived in \cite{Braun:2006td} taking into account the 
semi-disconnected pion-nucleon contributions in the intermediate state. We demonstrate that, with this addition, 
the applicability of the sum rules can be extended to the lower $Q^2$ region and the LET results in (\ref{LET}) are indeed
reproduced at $Q^2\sim 1$~GeV$^2$ to the required accuracy $\mathcal{O}(m_\pi)$. 
The results presented below essentially interpolate between the large-$Q^2$ limit considered in 
\cite{Braun:2006td} and the standard LET predictions at low momentum transfers. 

Accurate quantitative predictions are difficult for several reasons, e,g, because the nucleon distribution amplitudes
are poorly known. In order to minimise the dependence of various parameters in this work we only use 
the LCSRs to predict certain form factor ratios and then normalise to the electromagnetic nucleon form factors 
as measured in the experiment, see \cite{BIP2} for the details.
In particular we use the parametrisation of the proton magnetic form factor from \cite{Brash:2001qq} and for the 
neutron magnetic form factor from \cite{Bosted:1994tm}. For the proton electric form factor we 
use the fit \cite{Gayou:2001qd,Brash:2001qq} to the combined JLab data in the $0.5<Q^2<5.6$~GeV$^2$ 
range
\begin{eqnarray}
  \mu_p \frac{G_E^p}{G_M^p} = 1-0.13 (Q^2-0.04)
\label{GEp}
\end{eqnarray}
and put the neutron electric form factor to zero, which should be good to our accuracy.
Note that using (\ref{GEp}) for larger values of $Q^2$ up to 10 GeV$^2$ is an extrapolation
which may be not justified. 

\begin{figure}[t]
  \includegraphics[width=1.00\textwidth,angle=0]{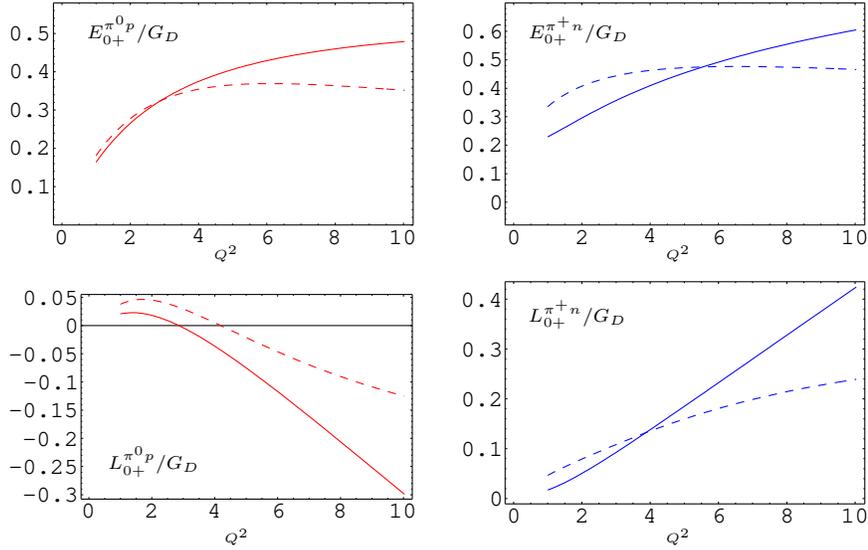}
\caption{The LCSR-based model (solid curves) for the $Q^2$ dependence of the electric and longitudinal 
partial waves at threshold $E_{0+}$ and $L_{0+}$, (\ref{waves}), in units of GeV$^{-1}$,  
normalised to the dipole formula (\ref{GD}). 
}
\label{fig:G12}
\end{figure}

The resulting LCSR-based model is shown by the solid curves in Fig.~\ref{fig:G12}, where the four partial waves at 
threshold that are related to the generalised form factors through the Eq.~(\ref{waves})  
are plotted as a function of $Q^2$, normalised to the dipole formula
\begin{equation}
   G_D(Q^2) = 1/(1+Q^2/\mu_0^2)^2
\label{GD}
\end{equation}   
where $\mu_0^2 = 0.71$ GeV$^2$.
This model is used in the numerical analysis presented below. We expect that its accuracy
is about 50\%. It can be improved in future by the calculation of radiative corrections
to the LCSRs, especially  if sufficiently accurate lattice calculations of the moments of nucleon 
distribution amplitudes become available.    
To give a rough idea about possible uncertainties, 
the ``pure'' LCSR predictions (all form factors and other input taken from the sum rules) 
are shown  by dashed curves for comparison.  

\section{Moving Away From Threshold} 

As a simple approximation, we suggest to calculate pion production near threshold 
in terms of the generalised form factors (\ref{def:G12}) and taking into account pion emission from the final state 
which dominates the P-wave contribution in the chiral limit (cf.\cite{PPS01}). 
In particular, we use the following expression:
\bea{def:amp}
 \lefteqn{\langle N(P')\pi(k) |j_\mu^{em}(0)| p(P)\rangle=}
\nonumber\\ &=& - \frac{i}{f_\pi} \bar N(P')\gamma_5
  \left\{\left(\gamma_\mu q^2 - q_\mu \!\not\! q\right) \frac{1}{m^2} G_1^{\pi N}(Q^2)
    - \frac{i \sigma_{\mu\nu}q^\nu}{2m} G_2^{\pi N}(Q^2)\right\}N(P)\,
\nonumber\\&&
{}+\frac{i c_\pi g_A}{2f_\pi[(P'+k)^2)-m^2]}\bar N(P')\not\! k \,\gamma_5(\not\!P'+m)
  \left\{F_1^p(Q^2)\left(\gamma_\mu-\frac{q_\mu\!\not\! q}{q^2}\right)\right.
\nonumber\\&&{}\left.
+ \frac{i\sigma_{\mu\nu} q^\nu}{2m}F_2^p(Q^2)\right\}N(P)\,.
\eea
Here $F_1^p(Q^2)$ and $F_2^p(Q^2)$ are the Dirac and Pauli electromagnetic form factors of the proton,
$c_{\pi^0} = 1$ and $c_{\pi^+} = \sqrt{2}$ is the isospin factor, $g_A=1.267$ and $f_\pi = 93$~MeV.

The separation of the generalised form factor contribution and the final state emission
in (\ref{def:amp}) can be justified in the chiral limit $m_\pi \to 0$ but involves ambiguities in contributions  $\sim \mathcal{O}(m_\pi)$.
We have chosen not to include the term $\sim \not\!k$ in the numerator of the proton propagator in the third line
in (\ref{def:amp}) so that this contribution strictly vanishes at the threshold. In addition, we found it convenient to include
the term $\sim q_\mu\!\not\! q /q^2 $ in the Lorentz structure that accompanies the $F_1$ form factor in order to make the
amplitude formally gauge invariant. To avoid misunderstanding, note that our expression is not suitable for making a
transition to the photoproduction limit $Q^2=0$ in which case, e.g. pion radiation from the initial state has to be taken
in the same approximation to maintain gauge invariance.

The virtual photon cross section can be written as a sum of terms
\begin{eqnarray}
 d\sigma_{\gamma^\ast} = \frac{\alpha_{\rm em}}{8\pi} \frac{k_f}{W} \frac{d\Omega_\pi}{W^2-m^2} |{\mathcal M}_{\gamma^*}|^2
\label{sigmagamma1}
\end{eqnarray}
with
\begin{eqnarray}
 |{\mathcal M}_{\gamma^*}|^2
&=& M_T
   + \epsilon \, M_L
   + \sqrt{2\epsilon(1+\epsilon)}\,M_{LT}\,\cos(\phi_\pi)
\nonumber\\&&{}
   + \epsilon M_{TT}\,\cos(2\phi_\pi)
   + \lambda \sqrt{2\epsilon(1-\epsilon)}\,M'_{LT}\, \sin(\phi_\pi)\,;
\end{eqnarray}
in the last term $\lambda$ is the beam helicity.

The complete expressions for the invariant functions are rather cumbersome but are simplified significantly 
in the chiral limit $m_\pi \to 0 $ and assuming $k_f ={\mathcal O}(m_\pi)$.  We obtain
\begin{eqnarray}
f_\pi^2 M_T &=&
\frac{4 \vec{k}_i^2 Q^2}{m^2} |G_1^{\pi N}|^2
 + \frac{c_\pi^2 g_A^2 \vec{k}_f^2 }{(W^2-m^2)^2} Q^2 m^2 G_M^2
\nonumber\\&&{}
   + \cos\theta \frac{c_\pi g_A |k_i| |k_f|}{W^2-m^2} 4 Q^2 G_M {\mathrm Re}\, G_1^{\pi N}\,,
\nonumber\\
f_\pi^2 M_L &=&
\vec{k}_i^2 |G_2^{\pi N}|^2 +
\frac{4 c_\pi^2 g_A^2 \vec{k}_f^2}{(W^2-m^2)^2} m^4_N G_E^2
\nonumber\\&&{}
   - \cos\theta \frac{c_\pi g_A |k_i| |k_f|}{W^2-m^2} 4 m^2 G_E {\mathrm Re}\, G_2^{\pi N} \,,
\nonumber\\
f_\pi^2 M_{LT} &=& -\sin\theta \frac{c_\pi g_A |k_i| |k_f|}{W^2-m^2} Qm
 \Big[ G_M {\mathrm Re}\,G_2^{\pi N} +  4 G_E {\mathrm Re}\,G_1^{\pi N}\Big]\,,
\nonumber\\
f_\pi^2 M_{TT} &=&0\,,
\nonumber\\
f_\pi^2 M'_{LT} &=& -\sin\theta \frac{c_\pi g_A |k_i| |k_f|}{W^2-m^2} Q m
 \Big[G_M {\mathrm Im}\,G_2^{\pi N} - 4 G_E {\mathrm Im}\,G_1^{\pi N}\Big]\,,
\end{eqnarray}
where $\vec{k}_i$ is the c.m.s. momentum in the initial state.
Note that the single spin asymmetry contribution $\sim M'_{LT}$ involves imaginary parts of the
generalised form factors that arise because of the final state interaction.
In our approximation $M_{TT}=0$ which is because we do not take into account the D- and higher 
partial waves. 
Consequently, the $\sim\cos(2\phi)$ contribution to the cross section is absent.

\begin{figure}[ht]
  \includegraphics[width=0.49\textwidth,angle=0]{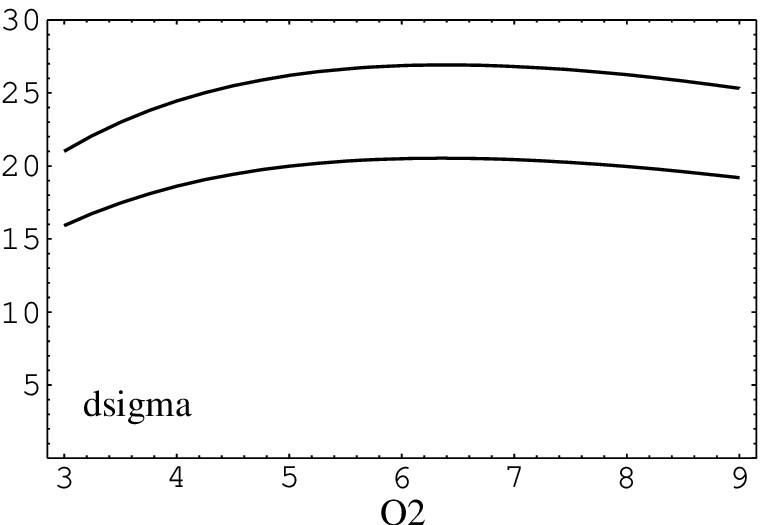}
  \includegraphics[width=0.49\textwidth,angle=0]{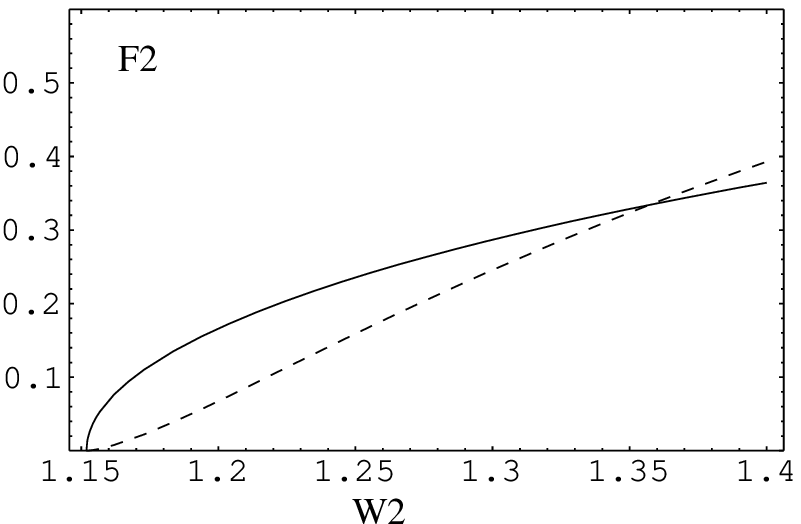}
\caption{{\it Left panel:}~ The integrated cross section $Q^6 \sigma_{\gamma^*p \to \pi^0 p}$ (in units of $\mu b$$\times$GeV$^6$)
as a function of $Q^2$ for $W=1.11$~GeV (lower curve) and $W=1.15$~GeV (upper curve).
 {\it Right panel:}~ The S-wave (solid) vs. the P-wave (dashed) contribution to the structure function $F^p_2(W,Q^2)$
as a function of $W^2$ for $Q^2 = 7.14$~GeV$^2$.
}
\label{fig:sigmagamma}
\end{figure}

We find that the integrated cross sections scale like $\sigma_{\gamma^*p \to \pi N} \sim 1/Q^6$, 
which is in agreement with the structure function measurements in the threshold region by E136 \cite{Bosted:1993cc}.
The S-wave contribution appears to be  larger than P-wave up to $W\simeq 1.16$~GeV.
The ratio of $\pi^0 p$ and $\pi^+ n$ final states is approximately $1:2$ and almost $Q^2$-independent.

\begin{figure}[ht]
  \includegraphics[width=0.49\textwidth,angle=0]{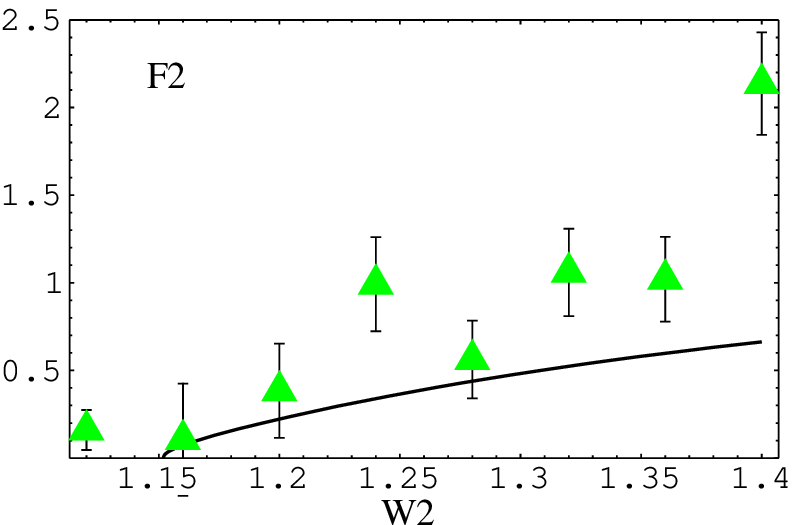}
  \includegraphics[width=0.49\textwidth,angle=0]{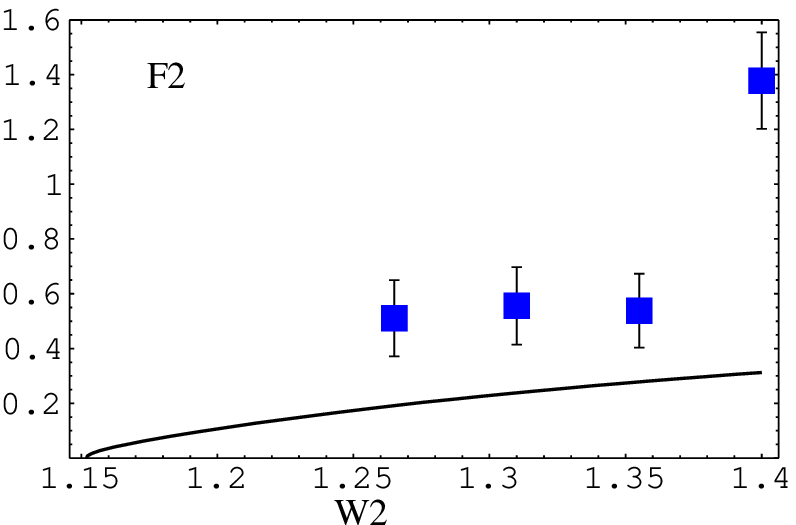}
\caption{The structure function $F^p_2(W,Q^2)$ as a function of $W^2$ scaled by a factor
 $10^3$ compared to the SLAC E136 data 
 at the average value $Q^2 = 7.14$~GeV$^2$ (left panel) and
$Q^2 = 9.43$~GeV$^2$ (right panel).
}
\label{fig:E136}
\end{figure}

The comparison of our calculation for the structure function $F^p_2(W,Q^2)$ in the threshold region $W^2<1.4$~GeV$^2$
to the SLAC E136 data \cite{Bosted:1993cc} at the average value $Q^2 = 7.14$~GeV$^2$ and 
$Q^2 = 9.43$~GeV$^2$ is shown in Fig.~\ref{fig:E136}.
The predictions are generally somewhat below these data ($\sim 50$\%), apart from the last data point at $W^2=1.4$~GeV$^2$
which is significantly higher.

Note that in our approximation there is no D-wave contribution, and the final state 
interaction is not included. Both effects can increase the cross section so that we consider the agreement as satisfactory.
We believe that the structure function at $W^2=1.4$~GeV$^2$ already contains a considerable D-wave contribution and also
one from the tail of the $\Delta$-resonance and thus cannot be compared with our model, at least in
its present form. 

To avoid misunderstanding we stress that the estimates of the cross sections presented here  
are not state-of-the-art and are only meant to provide one with the order-of-magnitude estimates
of the threshold cross sections that are to our opinion most interesting. 
These estimates can be improved in many ways, for example taking into account
the energy dependence of the generalised form factors generated by the FSI
and adding a model for the D-wave contributions. The model can also be tuned to reproduce the 
existing lower $Q^2$ and/or larger $W$ experimental data. 

\section*{Acknowledgements} 

We gratefully acknowledge discussions with A.~Afanasev, V.~Kubarovsky, A.~Lenz, A.~Sch{\"a}fer, P.~Stoler and
I.~Strakovsky on various aspects of this project. 
V.B. thanks U.~Meissner for bringing Ref.~\cite{Bernard:1992ys} to his attention and useful comments.
The work of D.I. was partially supported by grants from
RFBR-05-02-16211, NSh-5362.2006.2 and BMBF(06RY258).
The work by A.P. was supported by the Studienstiftung des deutschen Volkes.

\bibliographystyle{ws-procs9x6}

\end{document}